\documentclass[prd,aps,tightenlines,superscriptaddress,aps,showpacs]{revtex4}

\usepackage{amsmath}
\usepackage{verbatim}
\usepackage{graphicx}
\usepackage[usenames]{color}
\usepackage{psfrag}
\usepackage{epstopdf}
\usepackage{hyperref}
\def\beq{\begin{equation}}
\def\eeq{\end{equation}}
\def\bea{\setlength\arraycolsep{1.4pt}\begin{eqnarray}}
\def\eea{\end{eqnarray}}
\def\bit{\begin{itemize}}
\def\eit{\end{itemize}}

\begin{document}

\title{Updated constraints on the cosmic string tension}
\author{Richard Battye} \email{rbattye@jb.man.ac.uk}
\affiliation{Jodrell Bank Center for Astrophysics, University of Manchester, Manchester, M13 9PL  UK}
\author{Adam Moss} \email{adammoss@phas.ubc.ca}
\affiliation{ Department of Physics \& Astronomy, University of British Columbia, Vancouver, BC, V6T 1Z1  Canada}

\date{\today}

\begin{abstract}
We re-examine the constraints on the cosmic string tension from Cosmic Microwave Background (CMB) and matter power spectra, and also from limits on a stochastic background of gravitational waves provided by pulsar timing.  We discuss the different  approaches to modeling string evolution and radiation. In particular, we show that the unconnected segment model can describe CMB spectra expected from thin string (Nambu) and field theory (Abelian-Higgs) simulations using the computed values for the correlation length, rms string velocity and small-scale structure relevant to each variety of simulation. Applying the computed spectra in a fit to CMB and SDSS data we find that $G\mu/c^2< 2.6\times 10^{-7}$ ($2 \sigma$) if the Nambu simulations are correct and $G\mu /c^2< 6.4\times 10^{-7}$  in the Abelian-Higgs case. The degeneracy between $G\mu/c^2$ and the power spectrum slope $n_{\rm S}$ is substantially reduced from previous work. Inclusion of constraints on the baryon density from Big Bang Nucleosynthesis (BBN) imply that $n_{\rm S} <1$ at around the $4\sigma$ level for both the Nambu and Abelian-Higgs cases. As a by-product of our results, we find there is ``moderate-to-strong'' Bayesian evidence that the Harrison-Zel'dovich spectrum is excluded (odds ratio of $\sim 100:1$) by the combination of CMB, SDSS and BBN when compared to the standard 6 parameter fit.  Using the contribution to the gravitational wave background from radiation era loops as a conservative lower bound on the signal for specific values of $G\mu/c^2$ and loop production size, $\alpha$, we find that $G\mu /c^2< 7\times 10^{-7} $ for $\alpha c^2/ (\Gamma\, G\mu )\ll1$ and $ G\mu/c^2 < 5\times 10^{-11}/\alpha$ for $\alpha c^2/(\Gamma\, G\mu) \gg1$. 
\end{abstract}

\maketitle

%----------------- INTRODUCTION -----------------------

\section{Introduction} 

The idea that cosmic strings may have been a source of inhomogeneities in the Universe has been around for some time (see e.g. ref.~\cite{stringrev} for an overview). Even though it now seems that inflation is the primary source of the inhomogeneities which lead to anisotropies in the Cosmic Microwave Background (CMB), the idea has persisted. The main reason for this is that the amplitude of fluctuations produced by strings formed close to a Grand Unified Theory (GUT) phase transition ($\sim 10^{16}{\rm GeV}$) is of the order of observed density inhomogeneities on large scales, making it an attractive probe of physics at energies way above that possible in terrestrial accelerators.

It may seem very non-minimal to have two mechanisms creating inhomogeneities and contrary to the common-sense approach advocated by Occam's Razor. However, there are at least two classes of inflation models, linked to popular ideas in fundamental physics, where strings can be formed and  the two sets of perturbations have similar amplitudes. One is based on Supersymmetric Hybrid Inflation~\cite{hybrid1,hybrid2,fterm,dterm1,dterm2} and the other is Brane Inflation~\cite{Dvali:1998pa}. In both cases cosmic strings can be formed during the phase transition at the end of inflation~\cite{10pera,10perb,10perc}. It was shown in ref.\cite{Jeannerot:2003qv} that this can happen in a large number of symmetry breaking schemes and it is possible that semi-local strings could form~\cite{Urrestilla:2004eh,Urrestilla:2007sf} in situations where strings are not forced to exist on topological grounds.

One cannot rule out strings using observations per se, rather one can only constrain the tension (or the mass per unit length) of the strings, and hence the energy scale at which the strings are formed. The tension is usually quantified in terms of the dimensionless ratio $G\mu/c^2$.  This can be done in many different ways. In this paper we will focus on two: the anisotropies, created by the Kaiser-Stebbins effect~\cite{Kaiser:1984iv,Fraisse:2007nu} and related pre-recombination processes in the CMB; and the stochastic background of gravitational waves created by the decay of string loops~\cite{Vachaspati:1984gt}.

There are serious difficulties in setting such limits due to the dynamic range involved in modeling the evolution of a cosmic string network over cosmological scales. This has led to a number of different constraints being published in the literature which are often contradictory -- these will be reviewed in the subsequent sections. The objective of the present paper is two-fold. For both types of observations we will review the important physics and attempt to delineate the key uncertainties. We then present what we believe are the most reliable constraints based on the present knowledge. In the case of the CMB observations we will present two different constraints, based on different assumptions about the evolution of the string network. We note that the constraints presented here will apply only directly to strings where the intercommutation probability, $p$, is unity; that is, they can only be directly applied to normal field theoretic strings and not necessarily to cosmic superstrings~\cite{superstrings}.

%----------------- CMB SPECTRA -----------------------

\begin{figure}
\centering
\mbox{\resizebox{0.4\textwidth}{!}{\includegraphics[angle=0]{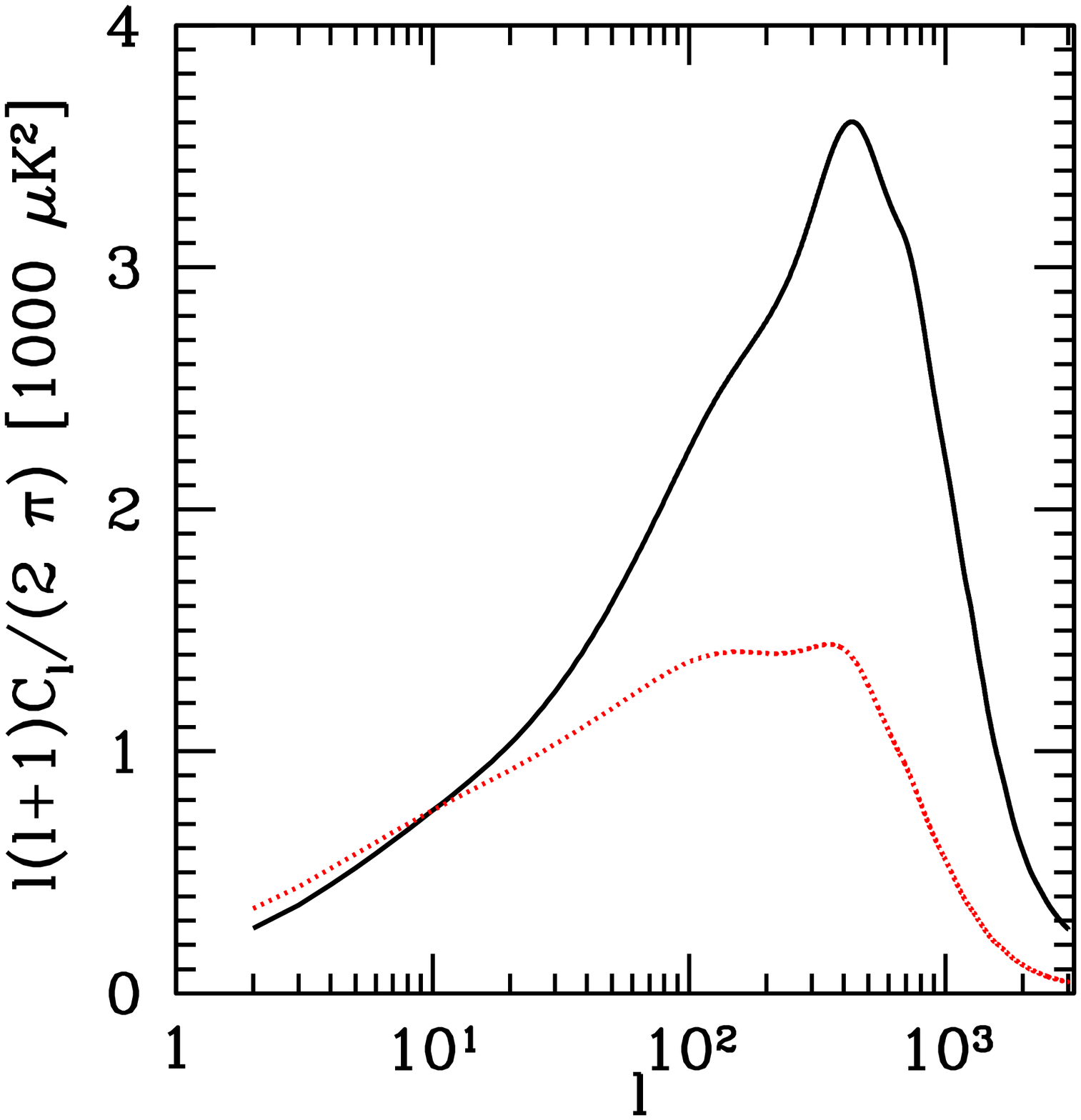}}}
\mbox{\resizebox{0.4\textwidth}{!}{\includegraphics[angle=0]{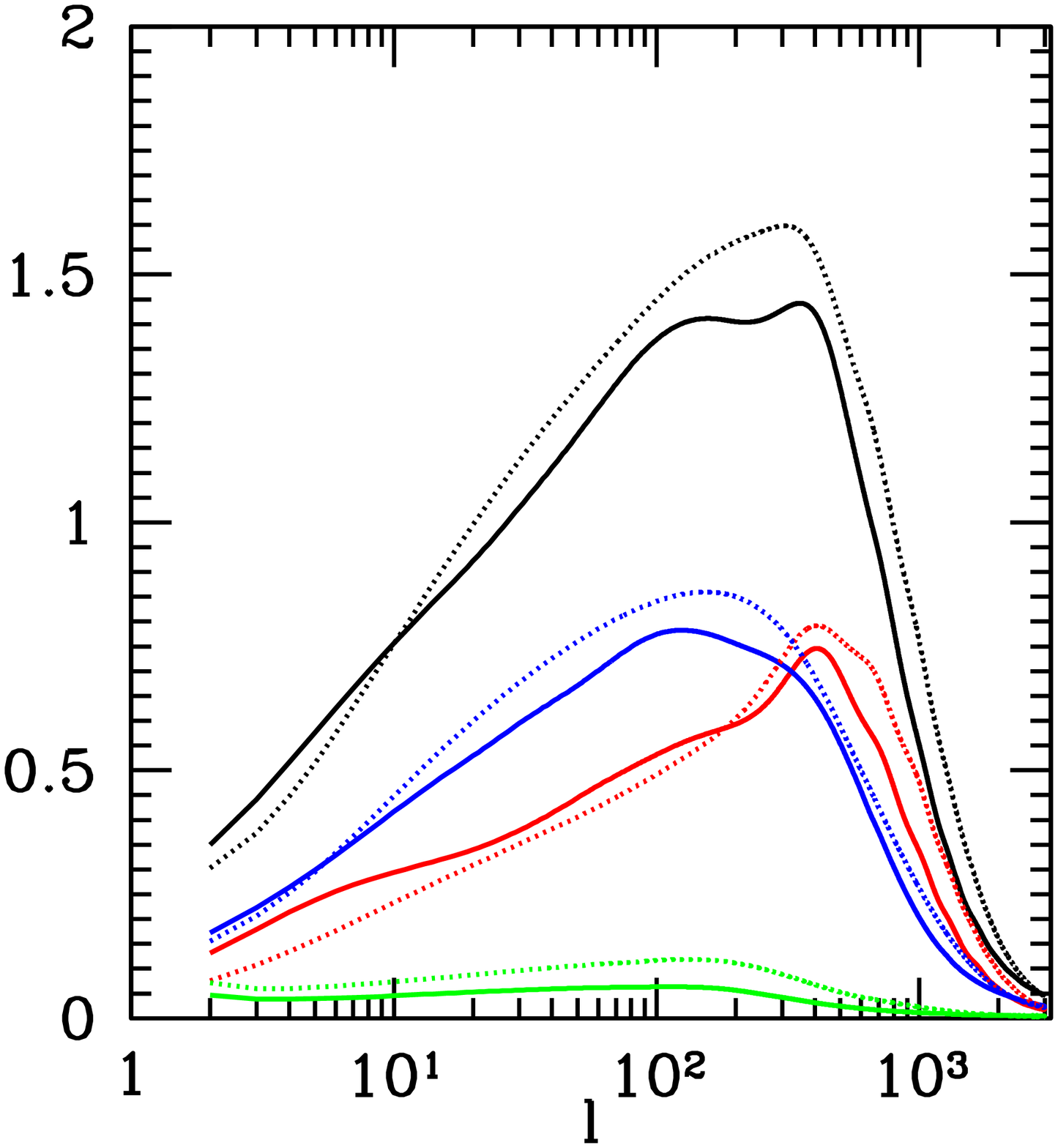}}}
\caption{\label{fig:comp} (Left) Comparison of cosmic string power spectra computed with the Unconnected Segment Model (USM) using  the Nambu -- model A -- (solid line) and AH mimic -- model B -- (dotted line) parameters. Spectra have been normalized to the WMAP value of $C_{10}$, giving $G\mu  / 10^{-6}c^2=1.18$ for the Nambu model and $1.91$ for the AH mimic. (Right) Comparison of scalar, vector and tensor modes for the USM AH mimic (solid) parameters and the actual AH spectra from simulations (dotted)~\cite{Bevis:2006mj}. The former has $G\mu / 10^{-6}c^2=1.91$, and the latter $2.04$. From top-to-bottom at low $\ell$ the ordering of the curves is the total power, then the anisotropy from vectors, scalars and tensors, respectively.}
\label{fig:spectra}
\end{figure}

\begin{figure}
\centering
\mbox{\resizebox{0.7\textwidth}{!}{\includegraphics[angle=0]{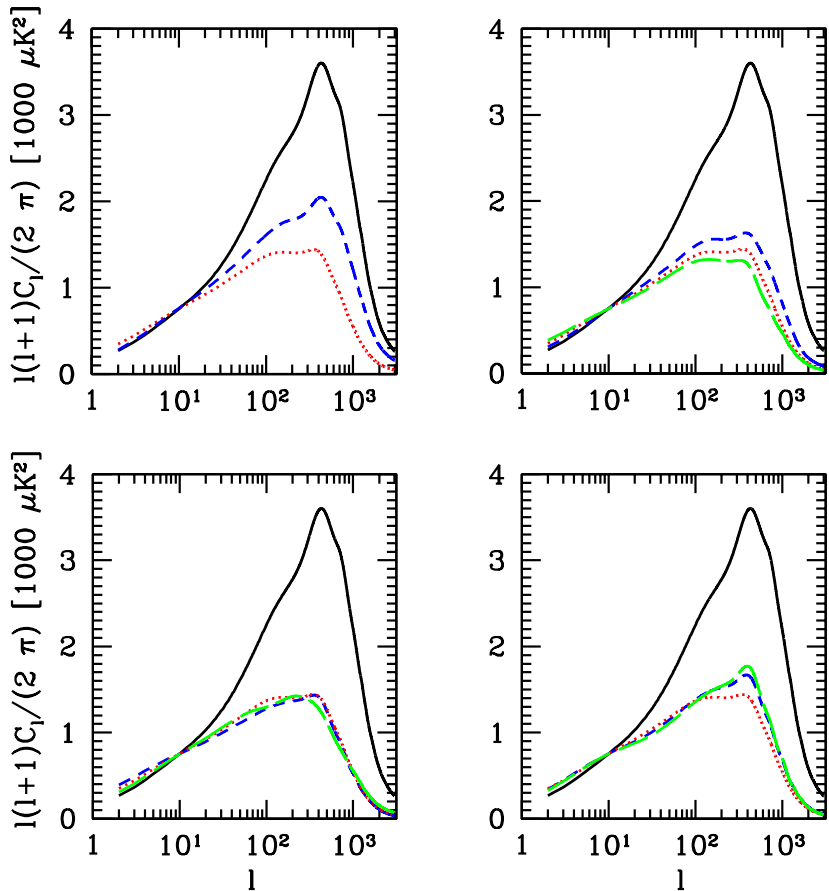}}}
\caption{A montage of plots which show the dependence of the string spectrum on parameters of the USM. In each case the solid and dotted lines are the Nambu and AH mimic models respectively from  Fig.~\ref{fig:spectra}. In all but the top-left plot we have varied the one parameter keeping all the others the same as in the AH mimic model: (top-left) the dashed line is the USM spectrum with $\xi=0.13$, $v=0.65c$ and $\beta=1.9$ but no matter-radiation transition; (top-right) the short dashed line has $\xi=0.2$ and the long dashed line has $\xi=0.5$; (bottom-left) the short dashed line has $v=0.0c$ and the long dashed line has $v=0.8c$; (bottom-right) the short dashed line has $\beta=1.5$ and the long dashed line $\beta=1.9$.}
\label{fig:morph}
\end{figure}

\section{CMB spectra from cosmic strings} \label{sec:spectra}

The key to predicting the observed CMB anisotropies from strings is a detailed understanding of the evolution of a cosmic string network. The main idea is that, after formation, the network evolves toward a self-similar scaling regime, whereby the average properties of the network remain approximately constant as a function of time. In the standard scenario this is achieved by the production of loops and their subsequent decay into radiation, usually assumed to be gravitational.

The scaling assumption is not at issue. It has been observed in just about every simulation that has ever been done. However, the precise characteristics of the scaling regime are of critical importance for predicting the CMB anisotropy spectrum. The problem is that each simulation technique needs to make assumptions in order to achieve sufficient dynamical range. The two most popular approaches are: (A) to solve the Nambu equations of motion for a string in an expanding universe and include the effects of reconnection by hand; and (B) to solve the equations of motion for the Abelian-Higgs (AH) model. In method A one ignores the effects of radiation backreaction, which is likely to be important in setting the size of loops which the string network creates, making the observed scaling dependent on the removal of sufficiently small loops from the simulation. Radiation into propagating modes of the field  is included in method B, but one is necessarily forced to consider simulation box sizes which are only $\sim 300$ times greater than the core string width, thereby reducing the dynamical range. This could also lead to the radiation being stronger in the simulation than in realistic string networks since the typical radius of curvature is large relative to the core width. Moreover one has to accept contortions involving fixing, or reducing the rate of growth of, the core width in comoving units in order to perform simulations in an expanding universe. Dynamical range is less of a problem in method A, since the core width is zero and the strings are represented as 1D objects, as opposed to the 3D box required for field theory simulations, thereby allowing much larger grid sizes. Having said that no simulation can achieve the dynamical range required to model the real Universe!

Simulations using method A have been used to derive the key statistical parameters of the string network which appear to be different in the radiation and matter eras~\cite{Bennett:1989yp,Martins:1995tg,Martins:1996jp,Shellard:1989yi}. It is found that the correlation length $L=\xi d_{\rm H}(t)$, where $d_{\rm H}(t)$ is the particle horizon distance, differs substantially between the radiation and matter eras, with $\xi_{\rm rad}=0.13$ and $\xi_{\rm mat}=0.21$ (see refs .~\cite{Battye:1997hu,Pogosian:1999np} and references therein), corresponding to a significantly lower string density ($\rho_{\rm str}=\mu/L^2$ where $\mu$ is the string tension - whose value is the main subject of the paper) in the matter era. There is substantial small-scale structure on the long string network. This can be quantified in terms of a renormalized string tension~\cite{Carter:1990nb}, $\mu_{\rm eff}=\beta\mu$,  and it is found that $\beta_{\rm rad}=1.9$ and $\beta_{\rm mat}=1.5$. On the other hand, the RMS velocity of the string segments only varies very slightly between the different eras, with $v_{\rm rad}=\langle v_{\rm rad}^2\rangle^{1/2}=0.65c$ and $v_{\rm mat}=\langle v_{\rm mat}^2\rangle^{1/2}=0.60c$. It has been shown that a velocity dependent one-scale model can be used to describe the evolution between the radiation and matter eras, and the extrapolation into the cosmological constant dominated era. This typically involves a long transient with the string density only achieving its matter era value around $z\sim 100$.

Contrasting results are found using method B~\cite{Bevis:2006mj}. Typically very small loops are created from the fragmentation of  horizon scale loops and most of the energy loss, required in order to maintain scaling, is achieved by radiation of the massive modes associated with the AH model. $\xi \approx 0.3$ and $v\approx 0.5c$~\cite{Hindmarsh:2008dw} are computed from the simulations, with little difference observed between simulations performed in the radiation and matter eras. The nature of the observed small-scale structure on the strings is different but is compatible with  $\beta\approx 1$. 

We note that Ringeval {\it et al}~\cite{Ringeval:2005kr} find that $\xi_{\rm rad}\approx 0.16$ and $\xi_{\rm mat}\approx 0.19$ using Nambu simulations which corresponds to a change in density between the radiation and matter densities of $\approx 1.4$ as opposed to $\approx 2.6$ which would be the case if $\xi_{\rm rad}\approx 0.13$ and $\xi_{\rm mat}\approx 0.21$. This will lead to results more similar to those which come from the AH model. In addition an attempt to compute the spectrum directly from Nambu simulations has recently being made~\cite{Landriau:2010cb}.

In both cases one can in principal compute  the unequal-time correlator (UETC), $C_{\mu\nu\rho\sigma}=\langle\theta_{\mu\nu}(\tau)\theta_{\mu\nu}(\tau^{\prime})\rangle$,  of the string network stress-energy tensor, $\theta_{\mu\nu}$, from the simulation. Using the stiff source approximation, one can then compute the CMB anisotropies by either decomposing the UETC in eigenmodes~\cite{Pen:1997ae}, or averaging realizations with the correct statistical properties. An alternative approach is to model the UETC using what has become known at the unconnected segment model (USM). This models the string network as an ensemble of string segments of length $\xi d_{\rm H}(t)$ with an RMS velocity $v$. This idea~\cite{Vincent:1996rb} was first used in the context of the CMB in refs.~\cite{Albrecht:1997nt,Albrecht:1997mz} and was later adapted to include the matter-radiation transition~\cite{Battye:1997hu} and the effects of small-scale structure on the network via the `wiggliness' parameter, $\beta$~\cite{Pogosian:1999np}. The weakness of this approach is that there is some freedom due to the parameters of the model, $\xi$, $v$ and $\beta$, which could be functions of time, but these can be and have been computed from simulations. 

The present work uses the USM and the code CMBACT~\cite{Pogosian:1999np} which was also used in refs.~\cite{Pogosian:2004ny,Wyman:2005tu,Pogosian:2006hg,Battye1,Pogosian:2008am}. The spectrum computed using the USM and the parameters from the Nambu simulations (model A)  are presented in the left hand panel of Fig.~\ref{fig:spectra}, along with the spectrum computed from the AH model simulations~\cite{Bevis:2006mj}. It is clear that the two spectra are very different, with model A having a fairly sharp peak at around $\ell\sim 500$, whereas the AH spectrum has a lower and broader peak, and falls off very sharply at high $\ell$. It is not necessary for the string spectrum to be known to the level of precision of the inflationary component, since it is constrained to be $<10\%$ of the total anisotropy. However, the discrepancy illustrated here is larger than is acceptable and it leads to important quantitative differences  when each of the spectra are applied to the data.

The spectrum computed using the USM is not unique, being a function of the parameters, and in particular the first CMB spectrum computed using it~\cite{Albrecht:1997nt} looks very similar to that computed from the Abelian-Higgs simulations. This first spectrum just used a single, constant value for $\xi$ and $v$, and ignored the effects of the matter-radiation transition and small-scale structure, that is, it had $\beta=1$. It seems plausible that it might be possible to compute an AH `mimic' model using the USM (model B). In order to investigate this we ran a suite of simulations of the USM with 3 parameters, $\xi$, $v$ and $\beta$, all assumed to be constant with time, and then minimized the residuals to the true AH spectrum taking into account the decomposition into the scalar, vector and tensor (SVT) modes, all of which are present in cosmic string spectra. We found that $\xi=0.35$, $v=0.4c$ and $\beta=1.05$ gave the best fit.  These are quite close to the values computed from the AH model simulations discussed earlier. The spectrum created using model B is compared to that from the AH model simulations in the right hand panel of Fig.~\ref{fig:spectra}. We see that there is remarkably good agreement between the spectra and that the values of $G\mu/c^2$ required to achieve the observed level of fluctuations are very similar. This suggests that the discrepancy in the spectra between the USM used to describe the Nambu simulations and that from the AH model simulations is mainly due to the computed correlation properties, and the differences in the simulation techniques, not the subsequent methods used to infer the CMB spectrum. We have also computed the CDM power spectrum for both string models, but these are subdominant compared to the inflationary component when normalized to the observed large angle CMB amplitude~\cite{Albrecht:1997nt,Battye:1997hu}. Since the string contribution is limited to  $<10\%$ of the total anisotropy, the CDM power spectrum from strings is completely subdominant. 

It is worth investigating what are the primary reasons for the differences between the Nambu and AH spectra. In Fig.~\ref{fig:morph} we have first removed the matter-radiation transition, making the spectrum just a function of single values of $\xi$, $v$ and $\beta$. We have then varied each of the three parameters relative to the AH mimic model. It is clear that the most significant reason for the discrepancy is the effect of the matter-radiation transition, which is present in the Nambu simulations but not observed in the AH simulations. The other important effect comes from the correlation length $\xi$, whereas the parameters $v$ and $\beta$ appear to have little effect on the overall spectrum. The small-scale structure parameter, $\beta$, does play an important role in regulating the relative contribution of the scalar, vector and tensor components; values of $\beta>1$ have an enhanced scalar component relative to the vector component. The tensor component is low in all cases.

In the next section, where we describe constraints on $G\mu/c^2$, we will present results for the two spectra (model A and B) -- the qualitative conclusions are similar, but there are important quantitative differences, particularly in the upper limit on $G\mu/c^2$ . It is instructive to try understand why there is such a significant discrepancy in the properties derived from the 2 different simulation methods which leads to these differences. 

The most worrying aspect of the AH simulations is their size relative to the core width. It could be that the absence of small-scale structure is related to the artificially strong massive radiation seen in the simulations. It is plausible that this could smooth out the small-scale structure on the long strings and remove the need for loop formation. The fact that the string width is kept constant in comoving coordinates, which makes it artificially large relative to the horizon, may be responsible for there being little difference between the scaling behaviours in the radiation and matter eras. The achilles heel of the Nambu simulations is the absence of radiation which could over-produce small-scale structure. Clearly more work needs to be done to resolve these discrepancies, but it is reassuring that both models can be represented with the necessary level of accuracy using the USM. An alternative to presenting results for the two separate models would be to allow $\xi$, $v$ and $\beta$ to vary within the USM and marginalized over them as nuisance parameters.

%----------------- CMB CONSTRAINTS -----------------------

\begin{table}
\begin{center}
\begin{tabular}{|c||c|c|c||c|c|c||} \hline
& \multicolumn{6}{|c|}{Model} \\ \hline
 Parameter & NS  					  	  &  S (NAMBU)  			&  S (AH)  	      	& HZ NS      			& HZ S (NAMBU)		& HZ S (AH) \\ \hline
$\Omega_{\rm b} h^2$ & $0.0231 \pm 0.0005$  & $0.0239 \pm 0.0008$  		& $0.0240 \pm 0.0008$ 	& $0.0243 \pm 0.0004$  	& $0.0253 \pm 0.0007$  	& $0.0253 \pm 0.0005$	 \\ \hline
$\Omega_{\rm c} h^2$ & $0.107 \pm 0.003$       & $0.107 \pm 0.004$ 		& $0.107 \pm 0.004$ 	& $0.105 \pm 0.003$ 	& $0.105 \pm 0.003$   	& $0.105 \pm 0.003$   \\ \hline
$\theta_{\rm A}$ & $1.042 \pm 0.002$                  & $1.043 \pm 0.002$  		& $1.044 \pm 0.003$ 	& $1.046 \pm 0.002$ 	& $1.047 \pm 0.002$    	& $1.047 \pm 0.002$  \\ \hline
$\tau_{\rm R}$ & $0.087 \pm 0.017$                     & $0.088 \pm 0.017$ 		& $0.089 \pm 0.017$ 	& $0.113 \pm 0.018$ 	& $0.111 \pm 0.018$   	& $0.101 \pm 0.017$ \\ \hline
$\log (10^{10} P_{\cal R})$    & $3.04 \pm 0.04$ & $3.01 \pm 0.04$ 			& $3.02 \pm 0.04$  		& $3.11 \pm 0.04$ 		& $3.07 \pm 0.04$    		& $3.04 \pm 0.04$   \\ \hline 
$n_{s} $ & $0.957 \pm 0.013$                                & $0.958 \pm 0.013$ 		& $0.970 \pm 0.016$ 	& 1 					& 1   					& 1  \\ \hline
$G \mu /10^{-7}c^2$ 			  	& -                 & $ <2.6$ 	 			& $ <6.4$  		          & - 					&  $1.9 \pm 0.5$ 	&  $5.3 \pm 1.0$  \\
	($f_{\rm 10	  }$) 			& -		  &	$(<0.044)$ 			&	$(<0.093)$			& -					&  $(0.027 \pm 0.013)$  & $( 0.075 \pm 0.025)$ 	\\ \hline
$h  $ & $0.74 \pm 0.02$                                           & $0.75 \pm 0.02$ 			& $0.75 \pm 0.02$ 		& $0.77 \pm 0.01$ 		& $0.79 \pm 0.02$    		& $0.78 \pm 0.02$   \\ \hline \hline
$- \log \mathcal{L}$ & 1440.8                                  & 1440.2 				& 1440.3  				& 1446.7 				& 1445.2   			& 1442.2   \\ \hline 
$+ {\rm BBN}$ & 1442.0 					  & 1442.0 				& 1442.1	 			& 1450.4 				& 1450.5  				& 1448.4    \\ \hline \hline
\end{tabular}
\caption{ \label{tab:sdss} CMB + SDSS constraints on string models using $q_{\rm str}$ as the parameter. Models denoted by S use strings and NS indicates that we have not used strings in the MCMC, while HZ denotes a Harrison-Zel'dovich spectrum (i.e. $n_{\rm s}=1$). NAMBU refers to the USM spectrum using the parameters derived from the Nambu simulations and AH to that of the AH mimic model. Upper limits on $G\mu/c^2$ and $f_{10}$ are $2\sigma$. The $\chi^2$ values of the models can be computed from $\chi^2=-2\log{\cal L}$. }
\end{center}

\end{table}
\begin{table}
\begin{center}
\begin{tabular}{|c|c|c|c|c|} \hline
& CMB & +SDSS & +BBN & +SDSS/BBN \\ \hline
NAMBU & 2.8 (0.054) & 2.6 (0.044) & 2.2 (0.032) & 2.2  (0.030) \\ \hline
AH  & 6.8 (0.11) & 6.4 (0.093) & 5.1 (0.057) & 5.0 (0.055) \\ \hline
\end{tabular}
\caption{ \label{tab:gmu_summary} $2\sigma$ upper limits on the string tension $G\mu/10^{-7}c^2$ and $f_{10}$ (bracketed) using $q_{\rm str}$ as the parameter.  }
\end{center}
\end{table}

\begin{figure}
\centering
\mbox{\resizebox{0.99\textwidth}{!}{\includegraphics[angle=0]{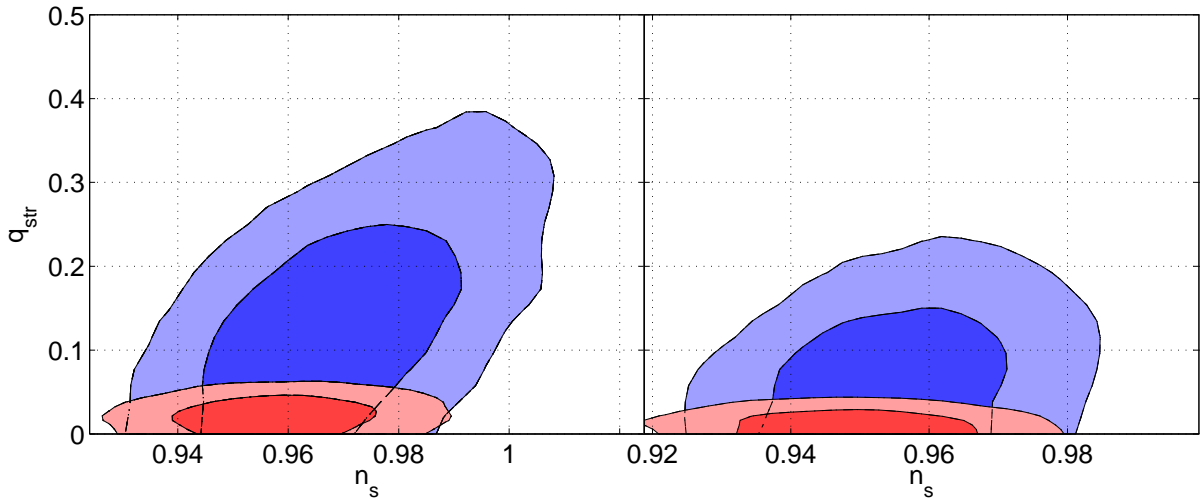}}}
\caption{\label{fig:gmu_ns} 2D likelihoods of scalar spectral index $n_{\rm S}$ versus cosmic string amplitude parameter $q_{\rm str}$ for the Nambu (red, likelihood surface at front) and AH (blue, surface behind) string spectra. In the left panel we show constraints from CMB+SDSS data, and on the right constraints from CMB+SDSS+BBN. }
\end{figure}

\section{Constraints on $G\mu/c^2$ from CMB anisotropies} \label{sec:constraints}

\subsection{Methodology}

Our methodology for obtaining CMB constraints on $G\mu/c^2$ will follow our previous work~\cite{Battye1} which focussed on the implications of the WMAP 3 year data release. Since the temperature power spectrum produced by strings is limited to around $\sim 10 \%$ of the total observed power, it is sufficient to keep the shape of the string spectrum fixed in a Markov-Chain-Monte-Carlo (MCMC) analysis, and vary only the normalization $G\mu/c^2$. In this way, one avoids the time-consuming computation of the string spectrum for each set of cosmological parameters. To generate the fixed string spectrum for the Nambu and AH mimic (hereafter we refer to this simply as `AH') models outlined above, we used best-fit parameters from the WMAP 5 year data release~\cite{Dunkley:2008ie}.

To generate the MCMC chains we used COSMOMC~\cite{cosmomc}, which makes use of the CAMB~\cite{Lewis:1999bs} and CMBFAST~\cite{Seljak:1996is} codes. We vary a total of 7 cosmological parameters: -- the baryon density, $\Omega_{\rm b}h^2$; cold dark matter density, $\Omega_{\rm c}h^2$; the acoustic scale, $\theta_{\rm A}$; optical depth, $\tau$; amplitude and spectral index of primordial fluctuations, $\mathcal {P_{\rm R}}$  and $n_{\rm S}$; together with the string normalization $G\mu/c^2$. We will include $G\mu/c^2$ in two ways. In one set of runs we adopt $q_{\rm str}=(G\mu/1.1\times 10^{-6}c^2)^2$ as the parameter. This was used in ref.~\cite{cbi}, and is similar to the method of ref.~\cite{bevis}, which used $f_{10}$, defined as the fraction of total power due to strings at $\ell=10$. An alternative parameterization is $\log_{10}(G\mu/c^2)$, which was used in ref.~\cite{Battye1}. As we will see there are subtle differences in the resulting constraints because the choices impose flat priors on different functions of $G\mu/c^2$.  In both approaches we will compute $f_{10}$ for each sample in the Markov-Chain  in order to compare with other analyses of string constraints~\cite{bevis}. Finally, we marginalize over the SZ amplitude $A_{\rm SZ}$, assuming the Komatsu and Seljak template~\cite{sztemplate}. Also, in some instances we will fix $n_{\rm S}=1$, that is a Harrison-Zel'dovich (HZ) spectrum, to compare the goodness-of-fit to a model where  $n_{\rm S}$ is allowed to vary.

We use CMB data from the WMAP 5 year release~\cite{Hinshaw:2008kr,wmap5} and the ACBAR~\cite{acbar}, BOOMERANG~\cite{boom}, CBI~\cite{cbi} and QUAD~\cite{quad} experiments, which observe to higher $\ell$. The latter high resolution data are important to help break the degeneracy which exists between $n_{\rm S}$ and $G\mu/c^2$ for WMAP alone. Following ref.~\cite{cbi}, we also include the BIMA~\cite{bima} data point at $\ell_{\rm eff} \approx 5000$ to constrain the SZ amplitude, without assuming a prior upper limit on $A_{\rm SZ}$ as done in the analysis of the WMAP 5 year data~\cite{wmap5}. However, imposing this upper limit on $A_{\rm SZ}$ does not significantly effect our conclusions. We also use matter power spectrum data from the SDSS Luminous Red Galaxies sample~\cite{sdss}, and Big Bang Nucleosynthesis (BBN) constraints on the baryon fraction from measurements of deuterium at high redshift~\cite{BBN}. The latter gives $\Omega_{\rm b}h^2=(2.13\pm 0.10)\times 10^{-2}$, which is slightly lower than that suggested from CMB data alone~\cite{Dunkley:2008ie}.

\subsection{Using $q_{\rm str}$ as a parameter}

The detailed results of our analyses using CMB+SDSS data and $q_{\rm str}$ as the parameter describing the amplitude of the string spectrum are given in table~\ref{tab:sdss}.  We include values for $-\log{\cal L}$ both with and without the BBN prior. We will use the results from CMB+SDSS data as our headline results in this paper, but will also discuss the significant impact of including BBN data.

For models where $n_{\rm S}$ is allowed to vary, the improvement in the goodness-of-fit when including strings is not significant over the non-string model, in both the Nambu and AH cases  ($\Delta\chi^2 = 1.2$ and $1.0$ respectively for one extra parameter in the case where we do not include the BBN prior). This suggests that there is no need to include the strings in order to explain the data, but constraints can be imposed on the cosmic string tension.

The constraints on $G\mu/c^2$ and $f_{10}$ are summarized for different combinations of data and the two string models in table~\ref{tab:gmu_summary}. It is clear that constraint is improved by the inclusion of the SDSS and BBN data. The $2\sigma$ upper limits which we will quote from CMB+SDSS data are $G\mu/c^2<2.6\times 10^{-7}$ for the Nambu model and $G\mu/c^2<6.4\times 10^{-7}$ for the AH model. The upper limit on $G\mu/c^2$ (and the respective string fraction $f_{10}$) is around a factor of two and a half smaller in the Nambu case and there are two reasons for this. The most significant is that the Nambu model spectrum has a lower value of $G\mu/c^2$ for a given power level on large-scales. In addition the Nambu spectrum has more small scale power relative to the AH case. This means that, if the amplitude of the spectrum is too high, the strings can modify the precise peak structure in the high WMAP signal-to-noise regime of  $\ell \approx 200-400$, leading to a tighter constraint.

The degeneracy between $G\mu/c^2$ and $n_{\rm S}$ is illustrated in the left panel of Fig.~\ref{fig:gmu_ns}. In previous work, we found that $n_{\rm S}=1$ was completely compatible with CMB data in the presence of a subdominant string contribution~\cite{Battye1}; this point was also made in ref.~\cite{bevis}. With improved CMB data and the inclusion of the SDSS, we find this is no longer the case, especially for the Nambu model. There is still a degeneracy between $G\mu$ and $n_{\rm S}$ in the AH model, although it is considerably reduced, but the two quantities appear to be completely decoupled in the Nambu case. The marginalized value of $n_{\rm S}$ is around $3\sigma$ from $n_{\rm S}=1$, with $\Delta\chi^2 = 10.0$ between the HZ spectrum and variable $n_{\rm S}$ models (for comparison, with no strings, $\Delta\chi^2 = 11.9$). For the AH case, the marginalized value of $n_{\rm S}$ is around $2\sigma$ from $n_{\rm S}=1$, with $\Delta\chi^2 = 3.8$. This point has led us to suggest that the F- and D-term supersymmetric hybrid inflation models  are very tightly constrained and that minimal models are probably ruled out by the data even in the presence of strings~\cite{BGMb}.

The values of $f_{10}$ we obtain from our analysis constrain the string component to be less than 4.4\% of the large scale amplitude in the case of the Nambu model and 9.3\% for the AH model, using the combination of CMB + SDSS data. In the AH case the values we find are  smaller than those quoted in ref.~\cite{bevis}. For CMB only, they found a nominal $2\sigma$ preference for strings with $f_{10} = 0.11 \pm 0.05$, whereas we only find a $2\sigma$ upper limit of $0.11$, corresponding to $G\mu/c^2 < 6.8\times 10^{-7}$. We presume that this is as a result of the improved CMB data.

It is noticeable that imposing the BBN prior reduces the upper limit on $G\mu/c^2$ and $f_{10}$.
This is because higher values of  $G\mu/c^2$ are degenerate with larger values of $\Omega_{\rm b} h^2$ and $n_{\rm S}$, as illustrated in ref.~\cite{Battye1}. This effect is more pronounced when using the AH model where there is still some degeneracy without the inclusion of the BBN prior. Applying the prior entirely removes the degeneracy between $G\mu/c^2$ and $n_{\rm S}$, as illustrated in the right panel of Fig.~\ref{fig:gmu_ns}. In this case, for {\em both} the Nambu and AH mimic models we find $n_{\rm S} <1$ at around $4\sigma$ significance ($\Delta\chi^2=16.8$ for the Nambu model and 12.8 for the AH model).

\subsection{Using $\log_{10} (G\mu/c^2)$ as a parameter}

\begin{figure}
\centering
\mbox{\resizebox{0.99\textwidth}{!}{\includegraphics[angle=0]{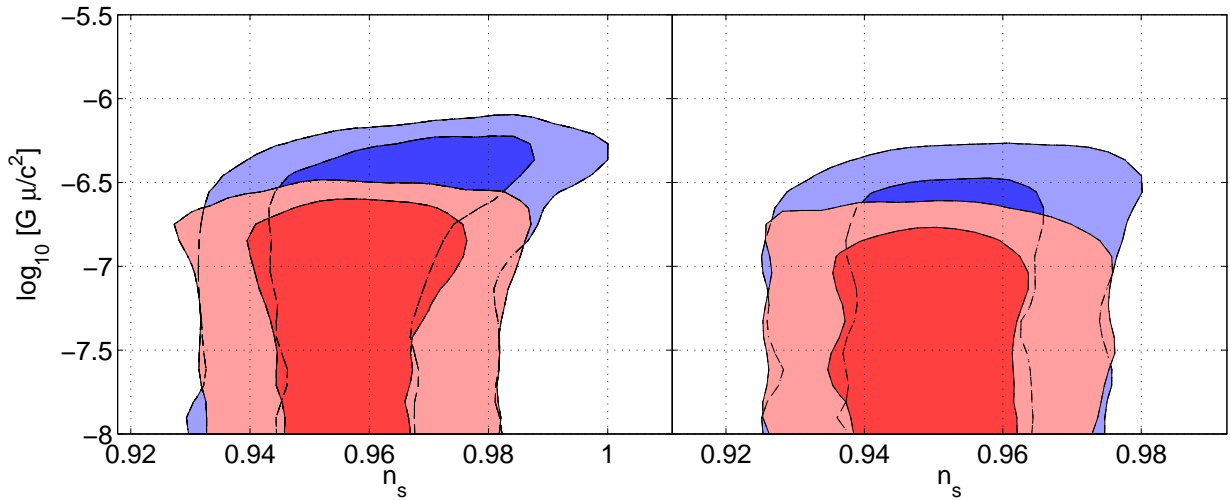}}}
\caption{\label{fig:loggmu_ns}  2D likelihoods of  $n_{\rm S}$ versus $\log_{10} [G \mu/c^2]$. Labeling is the same as in Fig.~\ref{fig:gmu_ns}.}
\end{figure}

For completeness we have included parameters and limits derived from using $\log_{10}(G\mu/c^2)$ as the parameter to describe the amplitude of the string power spectrum. The results are summarized in tables~\ref{tab:sdss_loggmu} and \ref{tab:loggmu_summary} along with Fig.~\ref{fig:loggmu_ns}. The results are broadly consistent with those obtained using $q_{\rm str}$ as the parameter, but there are some quantitative differences which come from the different prior space.  This means that there is a different measure used in the marginalization process. We note that $\log_{10}(G\mu/c^2)$ was used in our earlier work~\cite{Battye1}, whereas something closer to $q_{\rm str}$ was used in other previous work and this accounts for some of the discrepancies with the results reported in ref.~\cite{bevis}. There is, of course, no correct choice of parameter and therefore the differences represent the level of  ``natural uncertainty'' in the parameter estimation process.

In this case we find that the upper limits are modified to $G\mu/c^2<2.3\times 10^{-7}$ for the Nambu model and $<5.3\times 10^{-7}$ for the AH model using CMB+SDSS, and limits from other data combinations are also strengthened. There are also some small differences in the marginalized values of cosmological parameters which is a  result of the different weighting in the marginalization process. These differences are slightly larger for the AH model where there is more significant degeneracy between the cosmological parameters and $G\mu/c^2$. As expected, the values of $-\log{\cal L}$  are more or less unchanged between using $q_{\rm str}$  and $\log_{10} (G\mu/c^2)$.

\begin{table}
\begin{center}
\begin{tabular}{|c||c|c|c||c|c|c||} \hline
& \multicolumn{6}{|c|}{Model} \\ \hline
 Parameter & NS  					  	  &  S (NAMBU)  			&  S (AH)  	      	& HZ NS      			& HZ S (NAMBU)		& HZ S (AH) \\ \hline
$\Omega_{\rm b} h^2$ & $0.0231 \pm 0.0005$  & $0.0234 \pm 0.0007$  		& $0.0234 \pm 0.0007$ 	& $0.0243 \pm 0.0004$  	& $0.0248 \pm 0.0007$  	& $0.0251 \pm 0.0005$	 \\ \hline
$\Omega_{\rm c} h^2$ & $0.107 \pm 0.003$       & $0.107 \pm 0.004$ 		& $0.107 \pm 0.003$ 	& $0.105 \pm 0.003$ 	& $0.105 \pm 0.003$   	& $0.105 \pm 0.003$   \\ \hline
$\theta_{\rm A}$ & $1.042 \pm 0.002$                  & $1.043 \pm 0.002$  		& $1.043 \pm 0.002$ 	& $1.046 \pm 0.002$ 	& $1.046 \pm 0.002$    	& $1.046 \pm 0.002$  \\ \hline
$\tau_{\rm R}$ & $0.087 \pm 0.017$                     & $0.087 \pm 0.016$ 		& $0.088 \pm 0.017$ 	& $0.113 \pm 0.018$ 	& $0.112 \pm 0.019$   	& $0.103 \pm 0.018$ \\ \hline
$\log (10^{10} P_{\cal R})$    & $3.04 \pm 0.04$ & $3.02 \pm 0.04$ 			& $3.03 \pm 0.04$  		& $3.11 \pm 0.04$ 		& $3.09 \pm 0.04$    		& $3.05 \pm 0.04$   \\ \hline 
$n_{s} $ & $0.957 \pm 0.013$                                & $0.957 \pm 0.013$ 		& $0.960 \pm 0.014$ 	& 1 					& 1   					& 1  \\ \hline
$G \mu /10^{-7}c^2$ 			  	& -                 & $ <2.3$ 	 			& $ <5.3$  		          & - 					&  $< 2.5$  	&  $(4.7 \pm 1.4)$  \\
	($f_{\rm 10	  }$) 			& -		  &	$(<0.035)$ 			&	$(<0.071)$			& -				&$( <0.044)$	& $(0.062 \pm 0.028)$	\\ \hline
$h  $ & $0.74 \pm 0.02$                                           & $0.74 \pm 0.02$ 			& $0.74 \pm 0.02$ 		& $0.77 \pm 0.01$ 		& $0.79 \pm 0.02$    		& $0.78 \pm 0.02$   \\ \hline \hline
$- \log \mathcal{L}$ & 1440.8                                  & 1440.2 				& 1440.4  				& 1446.7 				& 1445.2   			& 1442.3   \\ \hline 
$+ {\rm BBN}$ & 1442.0 					  & 1442.1 				& 1442.0	 			& 1450.4 				& 1450.6  				& 1448.5    \\ \hline \hline
\end{tabular}
\caption{ \label{tab:sdss_loggmu} Equivalent of table~\ref{tab:sdss} but using $\log_{10} (G\mu/c^2)$ as the parameter defining the string amplitude.}
\end{center}
\end{table}

\begin{table}
\begin{center}
\begin{tabular}{|c|c|c|c|c|} \hline
& CMB & +SDSS & +BBN & +SDSS/BBN \\ \hline
NAMBU & 2.5 (0.042) & 2.3 (0.035) & 1.8 (0.021) & 1.8  (0.020) \\ \hline
AH  & 5.7 (0.084) & 5.3 (0.071) & 3.9 (0.038) & 3.8 (0.035) \\ \hline
\end{tabular}
\caption{ \label{tab:loggmu_summary} Equivalent of table~\ref{tab:gmu_summary} but using $\log_{10}(G\mu/c^2)$ as the parameter defining the string amplitude.}
\end{center}
\end{table}

\subsection{Bayesian evidence against the Harrison-Zel'dovich spectrum}

As a by-product of our analysis, we find there is strong evidence that $n_{\rm S} \ne 1$ within the 6 parameter family of $\Lambda$CDM models. With no strings, including the BBN prior increases the $\Delta\chi^2$ between HZ and variable $n_{\rm S}$ from $11.9$ to $16.8$ (as noted above, a similar $\Delta\chi^2$  for BBN is also true with strings). %This corresponds to the HZ model being excluded at  $4.1\sigma$ ($n_{\rm S} = 0.951 \pm 0.012$). 
One can further quantify this significance  by computing the Bayesian evidence using the Savage-Dickey method (see, for example, ref.~\cite{trotta}). To do this, we assume Gaussian prior information on $n_{\rm S}$ with mean 1 and standard deviation 0.2. The resulting Bayes factor between HZ and variable $n_{\rm S}$ models is  $\log B=-4.5$, corresponding to an odds ratio of $\sim 100:1$, which provides ``moderate-to-strong'' evidence that $n_{\rm S} \ne 1$. 

%----------------- PULSAR CONSTRAINTS -----------------------

\section{Constraints on $G\mu/c^2$ from Pulsar timing} \label{sec:pulsar}

\begin{figure}
\centering
\mbox{\resizebox{0.7\textwidth}{!}{\includegraphics[angle=0]{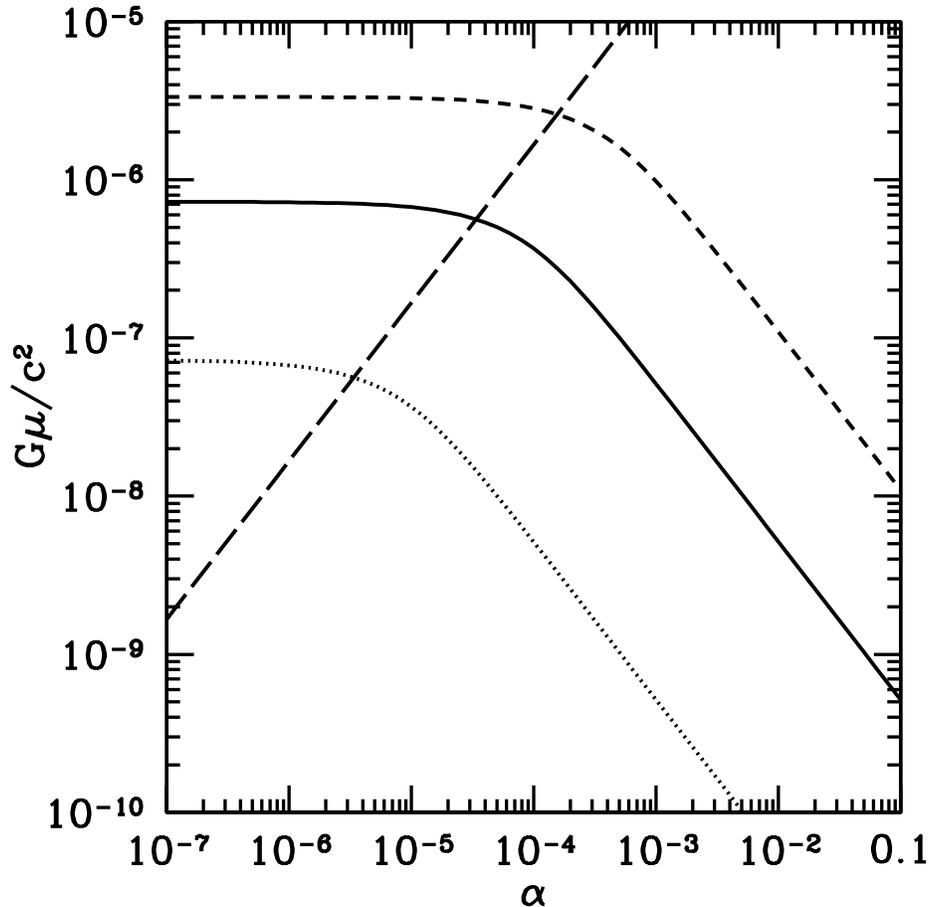}}}
\caption{\label{fig:pulsar} Pulsar constraints on  $G\mu/c^2$ as a function of dimensionless loop production size $\alpha$. The various lines show limits from different observational constraints on the gravitational wave background -- the solid line is derived from $\Omega_g h^2 < 2 \times 10^{-8}$~\cite{jenet}, the dotted line from $\Omega_g h^2 < 2 \times 10^{-9}$~\cite{vanlommen} and the short-dashed line from $\Omega_g h^2 < 9.3 \times 10^{-8}$~\cite{McHugh:1996hd}. We also show the relation $\alpha = 60 \, G\mu/c^2$ (long-dashed).}
\end{figure}

It is expected that the cosmic string network will emit gravitational waves, the dominant contribution coming from loop decay~\footnote{If the conclusions of the AH model simulations are to be taken at face value, then it could be that massive radiation is the dominant decay mechanism and gravitational wave production will be suppressed. If that were true then the constraints presented in this section would not apply, but there would be a strong constraint on $G\mu/c^2$ coming from the possible overproduction of high energy cosmic rays.}, leading to a stochastic background which can be constrained by pulsar timing and also recent measurements made by LIGO~\cite{Caldwell:1991jj,CBS,Damour:2000wa,Damour:2001bk,Damour:2004kw,Siemens:2006yp,DePies:2007bm}. At this stage those from pulsar timing are the strongest, and therefore we shall concentrate on them. Just as with the limits from the CMB anisotropy there are a number of constraints available in the literature, many of which are contradictory, mainly due to the different assumptions which are made.

There are two important issues here: the size of the loops produced by the string network and the spectrum of the radiation they produce. These issues require detailed modeling of the behaviour of the string network on small scales and an understanding of the impact of radiation backreaction.

The size of loops produced is usually assumed to be a constant fraction of the horizon size at formation, $\alpha=l(t_{\rm f})/d_{\rm H}(t)$, and will be related to the gravitational backreaction scale by $\alpha\sim\Gamma\, G\mu$, where $\Gamma\sim 60$ is defined below. The Nambu simulations do not include the effects of radiation backreaction, and we have argued that the AH simulations include massive radiation at an unnaturally strong level. Therefore, there is no reliable determination of the loop production size $\alpha$ from either type of simulation.

The spectrum of radiation emitted by given loop solutions of the Nambu equations of motion can be computed using the quadrupole formula under the assumption that backreaction does not effect the shape of the string trajectory. In particular one finds that the total power emitted by a loop is independent of the size the loop, $l$, and if given by $P=\Gamma \,G\mu^2c$. This can be written in terms of harmonics of half of the string length, $f_{n}=4\pi n/l$, as 
\begin{equation}
P=\sum_{n=1}^{\infty}P_n\,,
\end{equation}
where $P_{n}\propto n^{-q}$. Typically values of $q$ are 4/3 for loops which generate cusps~\cite{Vachaspati:1984gt}, a point on the loop which travels at the speed of light at some point during its evolution, and 2 for a square loop which has 4 kinks on it~\cite{Garfinkle:1987yw}. However, this simplification has a significant drawback: it assumes that the string solution is not affected by radiation backreaction. This would mean that the cusps and kinks, both of which are points on the string moving at the speed of light, are not rounded off by the radiation they produce.  

The issue of gravitational radiation backreaction is difficult to address quantitatively, but it might be possible to glean some understanding of the process by studying Goldstone boson radiation from global strings instead. Although it is typically much stronger,  this radiation is in many other ways similar to gravitational radiation, but in contrast it can be studied using a field theoretic approach.  It has been shown that kinks on  global strings are rapidly smoothed out by Goldstone boson radiation, and an initial kink perturbation is more or less indistinguishable from a sinusoidal perturbation after a few timesteps~\cite{Battye:1993jv}; a similar effect is expected for cusps. This suggests that the radiation spectrum from a loop will be modifed either by a change in the power law, or strong cut-off at large $n$.

In order to make contact with bounds on a stochastic background of gravitational waves from pulsar timing we need to estimate the spectral energy density of gravitational waves
\begin{equation}
\Omega_{\rm g}={f\over\rho_{\rm g}}{d\rho_{\rm g}\over df}\,,
\end{equation}
emitted by the network. This can be related to the dimensionless strain by
\begin{equation}
h_{\rm c}=1.3\times 10^{-20} \sqrt{\Omega_{\rm g}h^2}\left({100{\rm Hz}\over f}\right)\,.
\end{equation}

In ref.~\cite{CBS} the effects of varying the power spectrum of radiation from the strings were studied.  It was shown that the observed gravitational radiation background in the region relevant for pulsar observations ($f\sim 10^{-9}{\rm \, Hz})$ was sensitive spectrum of radiation emitted by the strings. In particular, it was shown that the amplitude of the background at these frequencies was around a factor of 3 higher when $q=4/3$ compared to $q=2$. A similar result was obtained by truncating the $q=4/3$ spectrum at some point $n_*\sim 1000$. The reason for this is that the emission at this particular frequency receives contributions from loops formed in the radiation era and also those from the matter era if they emit into high harmonics. Whatever one does with the spectrum of radiation emitted by loops from the matter era, the radiation era contribution will remain and, therefore,  given the uncertainties in the spectrum of radiation from loops discussed above, we will take a conservative approach and just use the radiation era contribution as a lower bound on the signal for a given $G\mu/c^2$. Tighter, but less reliable, constraints can be found by including the matter era contribution. 

We will assume that $\alpha$ is undetermined and use the analytic form for the radiation era spectrum presented in ref.~\cite{CBS}
\begin{equation}
\Omega_{\rm g}h^2=1.17\times 10^{-4}{G\mu\over c^2}\left({1-{\langle v_{\rm rad}^2\rangle/ c^2}\over \xi_{\rm rad}^2\Omega_{\rm m}}\right){(1+1.4x)^{3/2}-1\over x}\,,
\label{omg}
\end{equation}
where $x=\alpha c^2/(\Gamma \, G\mu)$ and $\Omega_{\rm m}$ is the total matter density relative to the critical density. In what follows we use parameters measured from the Nambu simulation to give $\xi_{\rm rad}$, $\langle v_{\rm rad}^2\rangle$ and we set $\Omega_{\rm m}=0.3$. We will ignore the uncertainties in these parameters since their cummulative effect is likely to be less than the uncertainties imposed by the lack of knowledge of $\alpha$ and the radiation spectrum. 

The constraints on $G\mu/c^2$ are presented in Fig.~\ref{fig:pulsar} for $\Gamma=60$ and as a function of $\alpha$ for a number of different possible values for the limit on $\Omega_{\rm g}h^2$. Probably the most reliable of the published limit is $\Omega_{\rm g}h^2<2\times 10^{-8}$~\cite{jenet}. There is stronger published limit of $2\times 10^{-9}$~\cite{vanlommen}, although  this is controversial~\cite{Damour:2004kw}. We have also included the value of $9.3\times 10^{-8}$~\cite{McHugh:1996hd} which is that used in ref.~\cite{CBS}. In each case the limit is flat for  $\alpha\ll\Gamma G\mu/c^2$ and is $\propto\alpha^{-1}$ for $\alpha\gg\Gamma G\mu/c^2$; this asymptotic behaviour can be understood easily from (\ref{omg}).  Using the limit of ref.~\cite{jenet} we find that  $G\mu /c^2< 7\times 10^{-7} $ for $\alpha c^2/ (\Gamma\, G\mu )\ll1$ and $ G\mu/c^2 <5\times  10^{-11}/\alpha$ for $\alpha c^2/(\Gamma\, G\mu) \gg1$.  In the former case this is weaker than the bound from the CMB and in the latter could be much stronger dependent on the value of $\alpha$. 

%----------------- CONCLUSIONS -----------------------

\section{Conclusions}

There are a wide range limits on $G\mu/c^2$ presented in the literature, often using identical or at the very least similar data. In this paper we have tried to clarify these issues and present what we believe are current constaints from the CMB and large scale structure data, and also from pulsar timing. We have presented evidence that the string spectra which come from different simulation techniques can all be modelled using the USM with parameters measured in simulations and deduce limits of $G\mu/c^2< 2.6\times 10^{-7}$ if the Nambu simulations are correct and $G\mu/c^2<6.4\times 10^{-7}$ if the AH simulations are correct. These use CMB+SDSS data and $q_{\rm str}$ as the parameter. We have also included data from BBN and used $\log[G\mu/c^2]$ as the parameter with similar but quantitatively different conclusions.

The limits from pulsar timing could already be much stronger than those from the CMB, for example, if $\alpha$ is large. However, the uncertainties in interpretation of the pulsar limits on $\Omega_{\rm g}h^2$ are much more severe. At this stage, it seems safest to use the low $\alpha$ limit of $\Omega_{\rm g}h^2<8\times 10^{-7}$,  which is one that is unlikely to have to be taken back.

\begin{figure}
\centering
\mbox{\resizebox{0.4\textwidth}{!}{\includegraphics[angle=0]{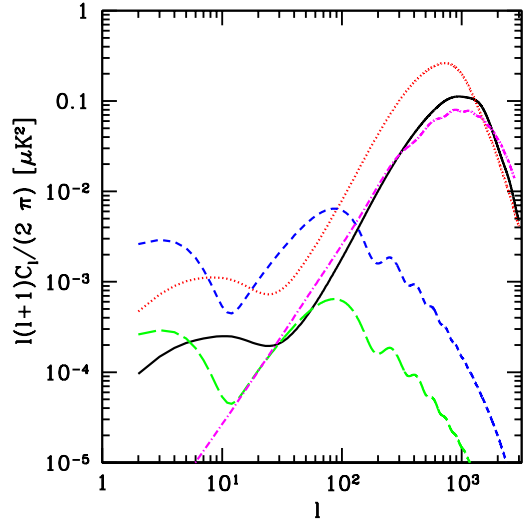}}}
\caption{\label{fig:bb} $B-$mode polarization power spectra from strings, tensors and gravitational lensing. For the string spectra, we use values of $G\mu/c^2$ corresponding to the $2\sigma$ upper limit from CMB+SDSS data, that is. $G\mu / c^2=2.6\times 10^{-7}$ for the USM Nambu model (solid) and $6.4\times 10^{-7}$ for the AH case (dotted). We also show the inflationary primordial tensor spectrum with $r=0.1$ at $k=0.05 \, {\rm Mpc}^{-1}$ (short dashed) and $r=0.01$ (long dashed). Finally, we show the gravitational lensing spectrum generated from $E-$mode mixing (dot-dash) expected in the inflationary model.}
\label{fig:bmode}
\end{figure}

As a final point we should note that a network of cosmic strings has another possible observational consequence in the CMB, albeit one of lower amplitude: the creation of B-mode polarization~\cite{Battye:1998js,Bevis:2007qz,Pogosian:2007gi} whose detection to very low values of $G\mu$ might be possible  if lense cleaning is found to be possible~\cite{Seljak:2006hi}. This is due to the fact that a substantial fraction of the anisotropies created by strings are vector modes. In Fig.~\ref{fig:bmode} we have presented the B-mode spectra for the Nambu and AH models with the value of $G\mu/c^2$ set to their repsective upper limits. These spectra are very different to those predicted in by inflationary models parameterized by the scalar-to-tensor ratio, $r$ which are also included in the figure. In fact, the spectra for $\ell >30$ are similar to  that produced by gravitational lensing particularly for tte Nambu case. This is because the B-mode polarization is created along the line of sight, albeit on different scales to the gravitational lensing pf the primary E-mode signal. These spectra comprise a white noise portion and a peak corresponding to the dominant scale which is on slightly larger scales in the AH model and, by coincidence, very similar to the lensing for the Nambu model. For the Nambu cas this  similarity will make them difficult to detect unless one can detect the analogue of the reionization bump which peaks around $\ell\approx 10$. It is, however, interesting since most inflationary models which produce strings create very low levels of gravitational waves. It has been shown that it is possible to discriminated them in the AH case~\cite{Urrestilla:2008jv}.

%----------------- ACKNOWLEDGMENTS -----------------------

\section*{Acknowledgments} This research was supported by the Natural Sciences and Engineering Research Council of Canada.  The calculations were performed on computing infrastructure purchased with funds from the Canadian Foundation for Innovation and the British Columbia Knowledge Development Fund. We thank Douglas Scott for useful discussions and comments on an earlier draft of the paper.


\begin{thebibliography}{99}
\newcommand{\prlet}{Phys.\ Rev.\ Lett.}
\newcommand{\npb}{Nucl.\ Phys.\ B}
\newcommand{\pletb}{Phys.\ Lett.\ B}
\newcommand{\prevd}{Phys.\ Rev.\ D}
\newcommand{\jhep}{J.\ High Energy Phys.}
\newcommand{\cqg}{Class.\ Quant.\ Grav.}
\newcommand{\jast}{Astrophys.\ J.}

\bibitem{stringrev}
A. Vilenkin and E. P. S. Shellard, {\em Cosmic Strings and Other Topological Defects} (2000), Cambridge University Press.

\bibitem{hybrid1}
 A.~D.~Linde, {\em Phys.\ Rev.\ D} {\bf 49} (1994) 748 [astro-ph/9307002]. 
 
 \bibitem{hybrid2}
 E.~J.~Copeland, A.~R.~Liddle, D.~H.~Lyth, E.~D.~Stewart and D.~Wands, {\em Phys.\ Rev.\ D} {\bf 49} (1994) 6410 [astro-ph/9401011].

\bibitem{fterm}
G.~R.~Dvali, Q.~Shafi and R.~K.~Schaefer, {\em Phys.\ Rev.\ Lett.}  {\bf 73} (1994) 1886 [hep-th/9406319].

\bibitem{dterm1}
P.~Binetruy and G.~R.~Dvali, {\em Phys.\ Lett.\ B} {\bf 388} (1996) 241 [hep-ph/9606342]. 

\bibitem{dterm2}
E.~Halyo, {\em Phys.\ Lett.\ B} {\bf 387} (1996) 43 [hep-th/9606423].

\bibitem{Dvali:1998pa}
G.~R.~Dvali and S.~H.~H.~Tye, {\em Phys.\ Lett.\ B} {\bf 450} (1999) 72 [hep-ph/9812483].

\bibitem{10pera}
R.~Jeannerot, Phys.\ Rev.\ D {\bf 56} (1997) 6205 [hep-ph/9706391]. 
 
 \bibitem{10perb}
C.~Contaldi, M.~Hindmarsh and J.~Magueijo, Phys.\ Rev.\ Lett.\  {\bf 82} (1999) 679 [astro-ph/9808201]. 

\bibitem{10perc}
R.~A.~Battye and J.~Weller, Phys.\ Rev.\ D {\bf 61} (2000) 043501 [astro-ph/9810203].

%\cite{Jeannerot:2003qv}
\bibitem{Jeannerot:2003qv}
  R.~Jeannerot, J.~Rocher and M.~Sakellariadou,
  %``How generic is cosmic string formation in SUSY GUTs,''
  Phys.\ Rev.\  D {\bf 68}, 103514 (2003)
  [hep-ph/0308134].
  %%CITATION = PHRVA,D68,103514;%%

%\cite{Urrestilla:2004eh}
\bibitem{Urrestilla:2004eh}
  J.~Urrestilla, A.~Achucarro and A.~C.~Davis,
  %``D-term inflation without cosmic strings,''
  Phys.\ Rev.\ Lett.\  {\bf 92}, 251302 (2004)
  [hep-th/0402032].
  %%CITATION = PRLTA,92,251302;%%

%\cite{Urrestilla:2007sf}
\bibitem{Urrestilla:2007sf}
  J.~Urrestilla, N.~Bevis, M.~Hindmarsh, M.~Kunz and A.~R.~Liddle,
  %``Cosmic microwave anisotropies from BPS semilocal strings,''
  JCAP {\bf 0807}, 010 (2008)
  [arXiv:0711.1842 [astro-ph]].
  %%CITATION = JCAPA,0807,010;%%

%\cite{Kaiser:1984iv}
\bibitem{Kaiser:1984iv}
  N.~Kaiser and A.~Stebbins,
  %``Microwave Anisotropy Due To Cosmic Strings,''
  Nature {\bf 310}, 391 (1984).
  %%CITATION = NATUA,310,391;%%

%\cite{Fraisse:2007nu}
\bibitem{Fraisse:2007nu}
  A.~A.~Fraisse, C.~Ringeval, D.~N.~Spergel and F.~R.~Bouchet,
  %``Small-Angle CMB Temperature Anisotropies Induced by Cosmic Strings,''
  Phys.\ Rev.\  D {\bf 78}, 043535 (2008)
  [arXiv:0708.1162 [astro-ph]].
  %%CITATION = PHRVA,D78,043535;%%

%\cite{Vachaspati:1984gt}
\bibitem{Vachaspati:1984gt}
  T.~Vachaspati and A.~Vilenkin,
  %``Gravitational Radiation From Cosmic Strings,''
  Phys.\ Rev.\  D {\bf 31}, 3052 (1985).
  %%CITATION = PHRVA,D31,3052;%%

\bibitem{superstrings}
J. Polchinski, (2004) [hep-th/0412244].

\bibitem{Bennett:1989yp}
D.~P.~Bennett and F.~R.~Bouchet,  {\em Phys.\ Rev.\ D} {\bf 41} (1990) 2408.

\bibitem{Martins:1995tg}
C.~J.~A.~P.~Martins and E.~P.~S.~Shellard, {\em Phys.\ Rev.\ D} {\bf 53} (1996) R575 [hep-ph/9507335]. 

\bibitem{Martins:1996jp}
C.~J.~A.~P.~Martins and E.~P.~S.~Shellard, {\em Phys.\ Rev.\ D} {\bf 54} (1996) 2535 [hep-ph/9602271].

\bibitem{Shellard:1989yi}
 E.~P.~S.~Shellard and B.~Allen, {\it  In ``Cambridge 1989, Proceedings, The formation and evolution of cosmic strings'' 421-448.}

\bibitem{Battye:1997hu}
R.~A.~Battye, J.~Robinson and A.~J.~Albrecht, {\em Phys.\ Rev.\ Lett.}  {\bf 80} (1998) 4847 [astro-ph/9711336].

\bibitem{Pogosian:1999np}
L.~Pogosian and T.~Vachaspati, {\em \prevd} {\bf 60} (1999) 083504 [astro-ph/9903361].

%\cite{Carter:1990nb}
\bibitem{Carter:1990nb}
  B.~Carter,
  %``Integrable equation of state for noisy cosmic string,''
  Phys.\ Rev.\  D {\bf 41}, 3869 (1990).
  %%CITATION = PHRVA,D41,3869;%%

\bibitem{Bevis:2006mj}
N.~Bevis, M.~Hindmarsh, M.~Kunz and J.~Urrestilla, {\em Phys.\ Rev.\ D} {\bf 75} (2007) 065015 [astro-ph/0606018].

\bibitem{Hindmarsh:2008dw}
M.~Hindmarsh, S.~Stuckey and N.~Bevis,
{\em Phys. Rev. D}{\bf 79} (2009) 123504 
[arXiv:0812.1929 [hep-th]].

%\cite{Ringeval:2005kr}
\bibitem{Ringeval:2005kr}
  C.~Ringeval, M.~Sakellariadou and F.~Bouchet,
  %``Cosmological evolution of cosmic string loops,''
  JCAP {\bf 0702}, 023 (2007)
  [arXiv:astro-ph/0511646].
  %%CITATION = JCAPA,0702,023;%%

%\cite{Landriau:2010cb}
\bibitem{Landriau:2010cb}
  M.~Landriau and E.~P.~S.~Shellard,
  %``Cosmic String Induced CMB Maps,''
  arXiv:1004.2885 [Unknown].
  %%CITATION = ARXIV:1004.2885;%%

%\cite{Pen:1997ae}
\bibitem{Pen:1997ae}
  U.~L.~Pen, U.~Seljak and N.~Turok,
  %``Power spectra in global defect theories of cosmic structure formation,''
  Phys.\ Rev.\ Lett.\  {\bf 79}, 1611 (1997)
  [arXiv:astro-ph/9704165].
  %%CITATION = PRLTA,79,1611;%%

%\cite{Vincent:1996rb}
\bibitem{Vincent:1996rb}
  G.~R.~Vincent, M.~Hindmarsh and M.~Sakellariadou,
  %``Scaling and small scale structure in cosmic string networks,''
  Phys.\ Rev.\  D {\bf 56}, 637 (1997)
  [astro-ph/9612135].
  %%CITATION = PHRVA,D56,637;%%

\bibitem{Albrecht:1997nt}
A.~J.~Albrecht, R.~A.~Battye and J.~Robinson, {\em Phys.\ Rev.\ Lett.}  {\bf 79} (1997) 4736 [astro-ph/9707129].

\bibitem{Albrecht:1997mz}
A.~J.~Albrecht, R.~A.~Battye and J.~Robinson,  {\em \prevd} {\bf 59} (1998) 023508 [astro-ph/9711121].

\bibitem{Pogosian:2004ny}
L.~Pogosian, M.~Wyman and I.~Wasserman, {\em JCAP} {\bf 0409} (2004) 008 [astro-ph/0403268].

\bibitem{Wyman:2005tu}
M.~Wyman, L.~Pogosian and I.~Wasserman, {\em \prevd} {\bf 72} (2005) 023513 [astro-ph/0503364].

\bibitem{Pogosian:2006hg}
L.~Pogosian, I.~Wasserman and M.~Wyman, [astro-ph/0604141].

\bibitem{Battye1} 
R.~A.~Battye, B.~Garbrecht and A.~Moss, {\em JCAP} 09 (2006) 007 [astro-ph/0607339].

\bibitem{Pogosian:2008am}
L.~Pogosian, S.~H.~H.~Tye, I.~Wasserman and M.~Wyman, {\em JCAP} 0902 (2009) 013 [arXiv:0804.0810 [astro-ph]].

\bibitem{Dunkley:2008ie}
J.~Dunkley {\em et al}, {\em Astrophys.\,J.\,Supp.} {\bf 180}, (2009) 306 [arXiv:0803.0586 [astro-ph]].

\bibitem{cosmomc}
A.~Lewis and S.~Bridle, {\em \prevd} {\bf 66} (2002) 103511 [astro-ph/0205436].

\bibitem{Lewis:1999bs}
A.~Lewis, A.~Challinor and A.~Lasenby, {\em \jast} {\bf 538} (2000) 473 [astro-ph/9911177].

\bibitem{Seljak:1996is}
U.~Seljak and M.~Zaldarriaga, {\em \jast} {\bf 469} (1996) 437 [astro-ph/9603033].

\bibitem{cbi}
J.~Sievers {\it et al.}, [arXiv:0901.4540 [astro-ph]]. 

\bibitem{bevis}
N.~Bevis, M.~Hindmarsh, M.~Kunz and J.~Urrestilla, {\em \prlet} {\bf 100} (2008) 021301 [astro-ph/0702223].

\bibitem{sztemplate}
E.~Komatsu and U.~Seljak, {\em MNRAS} {\bf 336} (2002) 1256 [astro-ph/0205468].

\bibitem{Hinshaw:2008kr}
G.~Hinshaw {\em et al}, {\em Astrophys.\,J.\,Supp.} {\bf 180} (2009) 225 [arXiv:0803.0732 [astro-ph]].

\bibitem{wmap5}
E.~Komatsu {\it et al.}, {\em Astrophys.\,J.\,Supp.} 180 (2009) 330 [arXiv:0803.0547 [astro-ph]].

\bibitem{acbar}
R.~Reichardt {\it et al.}, {\em \jast} {\bf 694} (2009) 1200 [arXiv:0801.1491 [astro-ph]]. 

\bibitem{boom}
W.~C.~Jones {\it et al.}, {\em \jast} {\bf 647} (2006) 823 [astro-ph/0507494].  

\bibitem{quad}
C.~Pryke {\it et al.}, {\em \jast} {\bf 692} (2009) 1247 [arXiv:0805.1944 [astro-ph]].

\bibitem{bima}
K.~Dawson {\it et al.}, {\em \jast} {\bf 647} (2006) 13 [astro-ph/0602413].

\bibitem{sdss}
M.~Tegmark {\it et al}, {\em Phys.\ Rev.\ D} {\bf 74} (2006) 123507 [astro-ph/0608632].

\bibitem{BBN}
M.~Pettini {\it et al}, [arXiv:0805.0594 [astro-ph]].

\bibitem{BGMb}
R.~A.~Battye, B.~Garbrecht and A.~Moss, [astro-ph/1001.0769]

\bibitem{trotta}
R.~Trotta, {\em MNRAS} 378 (2007) 72 [astro-ph/0504022].  


%\cite{Garfinkle:1987yw}
\bibitem{Garfinkle:1987yw}
  D.~Garfinkle and T.~Vachaspati,
  %``RADIATION FROM KINKY, CUSPLESS COSMIC LOOPS,''
  Phys.\ Rev.\  D {\bf 36}, 2229 (1987).
  %%CITATION = PHRVA,D36,2229;%%

%\cite{Caldwell:1991jj}
\bibitem{Caldwell:1991jj}
  R.~R.~Caldwell and B.~Allen,
  %``Cosmological Constraints On Cosmic String Gravitational Radiation,''
  Phys.\ Rev.\  D {\bf 45}, 3447 (1992).
  %%CITATION = PHRVA,D45,3447;%%

\bibitem{CBS}
R.~R.~Caldwell, R.~.A.~Battye and E.~.P.~S.~Shellard, {\em Phys.\ Rev.\ D} {\bf 54} (1996) 7146 [astro-ph/9607130].

%\cite{Damour:2000wa}
\bibitem{Damour:2000wa}
  T.~Damour and A.~Vilenkin,
  %``Gravitational wave bursts from cosmic strings,''
  Phys.\ Rev.\ Lett.\  {\bf 85}, 3761 (2000)
  [arXiv:gr-qc/0004075].
  %%CITATION = PRLTA,85,3761;%%

%\cite{Damour:2001bk}
\bibitem{Damour:2001bk}
  T.~Damour and A.~Vilenkin,
  %``Gravitational wave bursts from cusps and kinks on cosmic strings,''
  Phys.\ Rev.\  D {\bf 64}, 064008 (2001)
  [arXiv:gr-qc/0104026].
  %%CITATION = PHRVA,D64,064008;%%

%\cite{Damour:2004kw}
\bibitem{Damour:2004kw}
  T.~Damour and A.~Vilenkin,
  %``Gravitational radiation from cosmic (super)strings: Bursts, stochastic
  %background, and observational windows,''
  Phys.\ Rev.\  D {\bf 71}, 063510 (2005)
  [arXiv:hep-th/0410222].
  %%CITATION = PHRVA,D71,063510;%%

%\cite{Siemens:2006yp}
\bibitem{Siemens:2006yp}
  X.~Siemens, V.~Mandic and J.~Creighton,
  %``Gravitational wave stochastic background from cosmic (super)strings,''
  Phys.\ Rev.\ Lett.\  {\bf 98}, 111101 (2007)
  [arXiv:astro-ph/0610920].
  %%CITATION = PRLTA,98,111101;%%

%\cite{DePies:2007bm}
\bibitem{DePies:2007bm}
  M.~R.~DePies and C.~J.~Hogan,
  %``Stochastic Gravitational Wave Background from Light Cosmic Strings,''
  Phys.\ Rev.\  D {\bf 75}, 125006 (2007)
  [arXiv:astro-ph/0702335].
  %%CITATION = PHRVA,D75,125006;%%

%\cite{Battye:1993jv}
\bibitem{Battye:1993jv}
  R.~A.~Battye and E.~P.~S.~Shellard,
  %``Global string radiation,''
  Nucl.\ Phys.\  B {\bf 423}, 260 (1994)
  [arXiv:astro-ph/9311017].
  %%CITATION = NUPHA,B423,260;%%

\bibitem{jenet}
F.~A.~Jenet {\it et al.}, {\em \jast} {\bf 653} (2006) 1571 [astro-ph/0609013].  

\bibitem{vanlommen}
A.~N.~Lommen, [astro-ph/0208572].

%\cite{McHugh:1996hd}
\bibitem{McHugh:1996hd}
  M.~P.~McHugh, G.~Zalamansky, F.~Vernotte and E.~Lantz,
  %``Pulsar Timing And The Upper Limits On A Gravitational Wave Background: A
  %Bayesian Approach,''
  Phys.\ Rev.\  D {\bf 54}, 5993 (1996).
  %%CITATION = PHRVA,D54,5993;%%

%\cite{Battye:1998js}
\bibitem{Battye:1998js}
  R.~A.~Battye,
  %``Cosmic strings in a universe with non-critical matter density,''
  [astro-ph/9806115].
  %%CITATION = ASTRO-PH/9806115;%%

\bibitem{Pogosian:2007gi}
L.~Pogosian and M.~Wyman, {\em \prevd} {\bf 77} (2008) 083509 [arXiv:0711.0747].

%\cite{Bevis:2007qz}
\bibitem{Bevis:2007qz}
  N.~Bevis, M.~Hindmarsh, M.~Kunz and J.~Urrestilla,
  %``CMB polarization power spectra contributions from a network of cosmic
  %strings,''
  Phys.\ Rev.\  D {\bf 76}, 043005 (2007)
  [arXiv:0704.3800 [astro-ph]].
  %%CITATION = PHRVA,D76,043005;%%

%\cite{Seljak:2006hi}
\bibitem{Seljak:2006hi}
  U.~Seljak and A.~Slosar,
  %``B polarization of cosmic microwave background as a tracer of strings,''
  Phys.\ Rev.\  D {\bf 74}, 063523 (2006)
  [arXiv:astro-ph/0604143].
  %%CITATION = PHRVA,D74,063523;%%

%\cite{Urrestilla:2008jv}
\bibitem{Urrestilla:2008jv}
  J.~Urrestilla, P.~Mukherjee, A.~R.~Liddle, N.~Bevis, M.~Hindmarsh and M.~Kunz,
  %``On the degeneracy between primordial tensor modes and cosmic strings in
  %future CMB data from Planck,''
  Phys.\ Rev.\  D {\bf 77}, 123005 (2008)
  [arXiv:0803.2059 [astro-ph]].
  %%CITATION = PHRVA,D77,123005;%%

\end{thebibliography}
\end{document}